\documentclass[12pt]{article}
\usepackage[utf8]{inputenc} 
\usepackage{graphicx}
\usepackage{times}
\usepackage{sectsty}
\usepackage{titlesec}
\usepackage{float}
\usepackage{hyperref}
\usepackage{caption}
\usepackage{subcaption}
\usepackage{listings}
\usepackage{wrapfig}
\usepackage{cite}
\usepackage{xcolor, soul}
\sethlcolor{green}
\usepackage{array,ragged2e} 
\usepackage{booktabs}
\usepackage{indentfirst}
\allsectionsfont{\normalsize\bfseries}
\usepackage[letterpaper, margin=1in]{geometry}

\titlespacing\section{0pt}{12pt plus 2pt minus 0pt}{5pt plus 2pt minus 2pt} 
\titlespacing\subsection{0pt}{12pt plus 2pt minus 0pt}{5pt plus 2pt minus 2pt}

\usepackage{titling}
\title{\large \textbf{Engineering Students' Usage and Perceptions of GitHub Copilot in Open-Source Projects}}
\date{}
\author{Neha Rani, Jeevan Ram Munnangi, Austin Matthew Spangler, and Donald Honeycutt\\
\normalsize University of Florida, Gainesville, Florida, USA }

\begin{document}
\maketitle
\vspace{-20pt}
\section*{Abstract}
\noindent
The evolution of LLM has resulted in coding-focused models that are able to produce code snippets with high accuracy. More and more AI coding assistant tools are now available, leading to greater integration of AI coding assistants into integrated development environments (IDEs). 
These tools introduce new possibilities for enhancing software development workflows and changing programming processes.  
GitHub Copilot, a popular AI coding assistant, offers features including inline code autocompletion, comment-driven code generation, repository-aware suggestions, and a chat interface for code explanation and debugging. 
Different users use these tools differently due to differences in their perception, prior experience, and demographics. Furthermore, differences in feature use may affect users' programming process and skills, especially for programming learners such as computer science students.
While prior work has evaluated the performance of LLM-driven code generation tools, their use and usefulness for developers, especially computer science students, remain underexplored. 
Given that these tools are available to students, it's important to understand how they use them, as this will affect their learning and programming processes.
By analyzing how students interact with these tools, we aim to understand the implications of AI-assisted programming and the demographic and prior experience factors that impact its use and trust toward it. Understanding in this direction will inform the design of future educational tools that responsibly incorporate AI coding support for computing students. For our investigation, we conducted an exploratory survey-based study in which participants completed a survey after completing an open-source project issue using GitHub Copilot as part of a course. Study participants were engineering students enrolled in a software engineering course at a US university. 
Students were introduced to all the GitHub Copilot features, and they were free to use GitHub Copilot features as they seemed fit for the open-source project contribution. 
We analyzed students' use of each feature and their perceived usefulness. Further, we explore and analyze significant differences in GitHub Copilot usage and students' perceptions of it based on demographic factors. 
Our results show that students used the GitHub Copilot chat feature and code generation feature more than other features. Gender, programming proficiency, and familiarity with AI impacted the usage of the GitHub Copilot feature for assistance in completing the open-source project contribution.

\section{Introduction}
\noindent
The rapid evolution of large language models (LLMs) has led to systems capable of generating code at an incredibly exceeding rate. Various well-trained models have been developed that provide code generation and coding assistance with high accuracy \cite{wermelinger2023using}. For instance, Codex \footnote{https://openai.com/codex/}
, a 12‑billion‑parameter model trained on a corpus of 54 million GitHub repositories, has shown remarkable proficiency in code generation, achieving success on more than 70\% of the 164 Python problems with 100 samples \cite{vaithilingam2022expectation}.
Multiple AI coding assistants are available and integrated with IDEs such as GitHub Copilot, Junie by JetBrains \footnote{https://www.jetbrains.com/junie/}. 
Consequently, AI coding assistants have become commonly integrated into modern developer environments. Tools such as GitHub Copilot \footnote{https://github.com/features/Copilot} 
Claude Code \footnote{https://claude.com/product/claude-code} 
have integrated across companies, open source repositories and development workflows to provide developers with better context support \cite{Li2025AITeammates,dohmke2023economic}. These tools aim to support developers' work and reduce the time spent on routine tasks while also serving as an effective secondary resource to assist with code comprehension and debugging. 
Programming learners at intermediate and novice levels are now using AI coding assistants. Using AI coding assistants effectively for programming is considered a skill that university students should acquire \cite{10.1145/3641554.3701800}. Ease of availability makes AI coding assistants popular among computer science students. Use of such tools is observed in both formal and informal settings \cite{rani2025can}. However, how students use and perceive these tools remains unexplored. Developing understanding in this direction is important to understand the students' learning process. 

\bigbreak
\noindent
Although the benefits of such tools are substantial, their use by computing students in their learning phase raises concerns about the role of AI in learning programming and software development. 
Particularly in programming education, AI tools have been shown to alter student problem-solving approaches and shift effort away from syntax and language to understanding code behavior and applying concepts \cite{su15075614}. A shift from low-level details to higher-level skills and thinking has slowly been recognized. However, most of the current research focuses on introductory classes with specific tasks \cite{ghimire2024coding}. As a result, there is a clear lack of information on students' use of AI in real-life settings in terms of development and code contribution tasks. Further understanding of computing students' usage of AI coding assistants will be useful for programming instructors and other stakeholders.

\bigbreak
\noindent
GitHub Copilot is one of the most widely adopted AI coding assistants in both academic and industry environments \cite{dohmke2023economic}. With over 1 million developer activations, 20000 organizations assisted, and over a 30 \% acceptance of its suggestions per user, the widespread use of GitHub Copilot continues to grow \cite{dohmke2023economic}. Outside of accuracy, GitHub Copilot's large set of integrations, tool access, and ease of use allow GitHub Copilot to maintain a dominant product in AI coding assistants. These tools provide a range of features such as in-line code completion, comment-driven completions, context-aware suggestions, and a chat interface, allowing many channels of interaction, all capable of providing the same high level of support and accuracy. Prior research has identified common interaction patterns among developers, such as exploration, debugging, and iteration over a section \cite{10.1145/3643991.3645078}. These patterns 
emphasizes performance-based metrics does not consider how developers, or student developers, use, perceive, and interpret the AI assistance. Understanding students' trust in such an application will have a potential impact on its usage and how that usage evolves over time. Further, exploring how demographic and prior experience factors impact how computing students use these AI coding assistance tools will give a closer understanding of how programmers are being shaped. 

\bigbreak
\noindent
To better understand students' use of AI in the higher education computer science context, 
we explore the usage of AI coding assistants in a popular yet challenging task of making coding contributions to an open-source project 
\cite{10.1145/3641554.3701800}. Open source provides an environment for new developers with an interest in certain tools, applications, or games to collectively work to develop or update a software application. 
This environment is closest to an industry setting, as it requires knowledge and depth of multiple technologies to organize and carry out the steps needed to effectively contribute \cite{radermacher2014investigating}. 
Open-source provides students with an environment requiring navigating unfamiliar modules, understanding affected areas, and implementing changes. 
Open-source project has been identified as an opportunity for students to contribute to an existing code base as a way of providing them with experience of working on a real-life project and bridge the gap between the skills needed and what is taught \cite{begel2008struggles, craig2018listening}. The complex tasks making contributions to open-source projects increase the need for and use of AI assistance. This makes open source project contribution a suitable task for exploring the use of an AI coding assistant and its perception. Further, there is a lack of investigation into gender-based differences in the use of AI coding assistants, even though prior work has shown gender-based differences in AI awareness and AI adoption \cite{iddrisu2025gender, cachero2025gender}. 
 \bigbreak
\noindent
 Through our work, we aim to answer the following research questions:
 
 \textbf{RQ1: What are the GitHub features that are used by student programmers? }
 
 \textbf{RQ2: What is the level of perceived usefulness of each of the GitHub Copilot features as reported by students? }
 
 \textbf{RQ3: Do demographic factors such as gender, prior familiarity, and programming proficiency have an impact on the usage and perception of different GitHub Copilot features? }

\bigbreak
\noindent
To answer the research questions, we conducted a quasi-experimental survey-based study paired with a course open-source programming assignment. The study was conducted as part of a software engineering course at a large public university in the US. Students were given an initial introduction to GitHub Copilot features, then were free to use any feature to aid in the open-source contribution assignment.
Our investigation aims at understanding the level of usage of different GitHub Copilot features and the preference for one feature over another. Further, we explore students' perceptions of the usefulness of each feature identified during the coding contribution task. We also assess students' overall trust in AI coding assistants after using them. Additionally, understanding the differences in the student population, such as gender and familiarity, leads to varying usage and perception outcomes.  


\section{Background and Related Work}
\noindent
Modern software development is increasingly intertwined with AI coding assistants. Tools like GitHub Copilot 
have emerged as popular “AI pair programmers” that integrate into developers’ IDEs and workflows. Their adoption has been rapid: according to the 2025 Stack Overflow developer survey, 51\% of professional developers report using AI tools daily in their development process  \cite{stackoverflow2025survey}. Universities and educational environments are also seeing a steep increase in AI usage, a jump from 53\% in 2024 to 88\% in 2025 \cite{StudentGenerativeAISurvey2025_2025}. The adoption of AI tools is driven not only by convenience but also by continual improvements in the scale and accuracy of the LLMs that power them. 
The usage of these tools has been shown to improve developer output and efficiency across a variety of coding tasks \cite{imai2022github} \cite{zhang2023practices}. Notably, environments that use AI coding assistant tools—particularly GitHub Copilot—are shown to significantly decrease review time and increase the likelihood of merge acceptance \cite{xiao2024generative}. Many developers already use AI coding assistants. In a recent investigation, Song, Agarwal, and Wen \cite{song2024impact} used GitHub usage data and causal inference methods to show that GitHub Copilot adoption can raise project-level code contributions by around ~6\%(5.9 - 6.5\% depending on specification/version), while also increasing coordination effort among collaborators, particularly for core contributors. 
Developing efficient usage of an AI coding assistant is becoming a skill set for undergraduate students who will soon join the developer workforce. The first step in setting appropriate grounds for training undergraduate students to use an AI coding assistant is to understand their current usage. Then, understanding their pain points, needs, concerns, factors that impact usage, perceived trust in the AI coding assistant, and their perceived usefulness of the system in the assistance it provides with the coding project is essential. This helps in guiding the development of pedagogical standards for AI education and training. However, fine-grained data about how programming learners (computer science students) employ these tools, specifically, the scope, usage patterns, and factors affecting the usage of AI, remain scarce \cite{xiao2025self}. 

\bigbreak
\noindent
The majority of prior AI coding assistant research remains focused on task-level outcomes such as efficiency, clarity, and output \cite{song2024impact}. Zhou et al. \cite{zhou2025exploring}, through an analysis of GitHub and Stack Overflow data, evaluated user-reported issues with GitHub Copilot.
Some of the recent explorations have begun to explore students' use of an AI coding assistant \cite{stray2025developer, 10.1145/3641554.3701800}.
Shihab et al. \cite{shihab2025effects} provided 10 undergraduate computer science students with development tasks with and without GitHub Copilot. Results suggested that the tasks were completed more efficiently with the help of GenAI, with reduced completion time and improved test pass rates. More studies indicated similar increases in output and efficiency \cite{ziegler2022productivity}, and noted an additional increase in deleted code \cite{imai2022github}. Sandoval et al. \cite{sandoval2023lost} in their research asked university students to solve coding problems using a low-level programming language(C). Their results showed that LLM assistance produced bugs at the same rate as humans, while also corroborating productivity benefits. Asare et al. \cite{asare2023github} found that GitHub Copilot’s propensity to introduce security vulnerabilities was context-dependent, that is, it varied according to the coding scenario. 
 Recent work has inspected students' interactions with AI, a Case Study of CS1 Students \cite{amoozadeh2024student} shows early empirical evidence of how novice programmers engage with generative AI during problem solving.
 Students with lower performance tend to rely more heavily on AI assistance, indicating a potential disparity in the benefits of AI use \cite{margulieux2024self}. Additional studies highlight the need to carefully monitor student interactions with AI to mitigate potential drawbacks and educational impacts \cite{shihab2025effects} \cite{fan2025impact}.

\bigbreak
\noindent
 More closely rated recent work explores the use of GitHub Copilot in large open-source contribution projects by upper-division university students \cite{10.1145/3641554.3701800}. They found that students employed a one-shot prompting strategy to get the AI coding assistant to solve the entire programming problem at once and then followed up with more prompts to get to the code they desired. They also found that students trusted the code chat feature more than the code generation feature. They also used the GitHub Copilot chat feature the most among all other features.
 However, this closely related prior work does not explore how useful students found these features, their usage, and how the perception of trust, AI usage, and its usefulness were impacted by demographic factors such as gender or prior familiarity. 

\section{Study Details}
\begin{figure}[t]
\centering
\includegraphics[width=0.5\columnwidth]{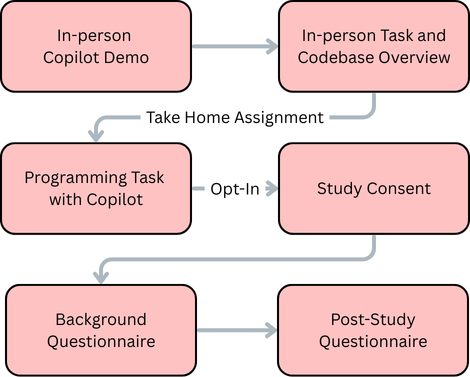} 
\caption{An overview of the study procedure.}
\label{study protocol}
\end{figure}

\noindent
The study was a quasi-experimental survey-based study.
The study was conducted with intermediate programmers (students) enrolled in the Software Engineering course at a large, public, research-intensive university. The study was approved by the university's IRB.
Participants were tasked with completing a programming assignment that involved adding a feature to an existing open-source codebase. 
Participants were required to complete the programming task using the Visual Studio Code IDE and were instructed to use the built-in GitHub Copilot integrations\footnote{https://code.visualstudio.com/docs/Copilot/overview} as they saw fit to help them complete the task.
To ensure basic familiarity with the GitHub Copilot features that are integrated into Visual Studio Code, participants were given an in-person demo of its core features and how to use them. 
The features they were shown are as follows:
\begin{itemize}
    \item Code autofill---suggests completions to partially written code, may be manually triggered or occur automatically when typing.
    \item Code generation---generates new code based on natural language prompts or in-line comments describing a desired function.
     \item Copilot chat---dedicated natural language chat that uses the code as additional context.
    \item Code explanation (/explain)---provides a natural language explanation of what a selected block of code does.
    \item Error fixing (/fix)---scans code and suggests fixes to any identified bugs or vulnerabilities.
    \item Documentation generation (/doc)---generates comments to document a selected portion of code.
\end{itemize}
\bigbreak
\noindent
Participants were provided the open source code for an interactive JSON visualization tool called JSON Crack\footnote{https://jsoncrack.com/}.
The system presents a graph visualization of nested JSON, and the programming task was to add node editing functionality within the visualization.
This task was designed to be of a reasonable level of difficulty for student programmers, yet sufficiently complex that participants would not trivially complete it on their own.
Because participants were adding to existing and unfamiliar code, all of the 
Participants had one week to complete the task as a take-home open-source contribution assignment. After completing the assignment, students participated in a opt in survey for the study.
Due to certain constraints on collecting students' grade information, we limited the study data to the survey. Students' performance in the assigned task was not collected or analyzed for the study
The survey included questions about demographics and prior experience with programming and AI tools. Participants were asked how much they used each of the GitHub Copilot features during the task and how useful they found them on a five-point Likert scale.
They were also asked to rate their agreement with several statements about GitHub Copilot from a series of human-machine trust scales developed by Madsen and Gregor~\cite{madsen2000measuring}.
The statements were grouped into five different trust constructs: reliability, understandability, technical competence, faith, and personal attachment.
The full list of statements can be found in Appendix \ref{sec:trustAppendix}.
The study procedure is illustrated in the figure \ref{study protocol}.

\section{Participants}
\noindent
A total of 189 students participated in the study. These were the students enrolled in the course who provided their consent to participate in the study. 
There were 10 incomplete responses. After removing incomplete responses, there were 179 complete responses. We analyzed the data from these 179 participants.
Out of these 179 participants, there were 130 male students, 47 female students, and 2 others.
The participant pool included 21 students in year 4 or more in their degree program, 103 in year 3, 52 in year 2, and 2 in year 1.
The participant pool consisted of the majority computer science major program. There were 119 students from the Computer Science program, 28 from the Computer Engineering program, 26 from the College of Liberal Arts \& Sciences, and 5 others.

\section{Data Analysis}
\noindent
To understand the prior experience of the participants, we first ran descriptive statistics for their prior familiarity with C++, LLMs, AI coding assistant, and their programming proficiency.
To understand which features of GitHub Copilot were used and how useful students found it in helping them complete the open-source project issue, we calculated the descriptive statistics of usage of each feature and its usefulness as reported in the survey response.
Further, to find if there are any significant differences between male and female usage and perceived usefulness of the GitHub Copilot features. We ran an independent T-test. Nonparametric alternatives were used if the data failed the normality test. The Shapiro-Wilk test was used to test the normality of the data. 
We also ran an independent T-test to examine any significant differences in the Madsen-Gregor Human-computer trust scale \cite{madsen2000measuring} and its sub-constructs based on gender.
Further, to determine whether participants' prior experience and expertise affect their use of GitHub Copilot features and their perceived usefulness, we conducted chi-square tests. This resulted in multiple chi-square tests, such as the impact of programming proficiency on use of individual GitHub Copilot features, the impact of programming proficiency on trust, the impact of familiarity with C++ on use of individual GitHub Copilot features, and the impact of familiarity with C++ on trust, the impact of familiarity with AI coding assistant on use of individual GitHub Copilot features, and the impact of familiarity with AI coding assistant on trust. 
The test results are reported in the results section.


\section{Results}
\noindent
\subsection{Prior familiarity with Programming and AI}
\begin{figure}[t]
\centering
\includegraphics[width=0.5\columnwidth]{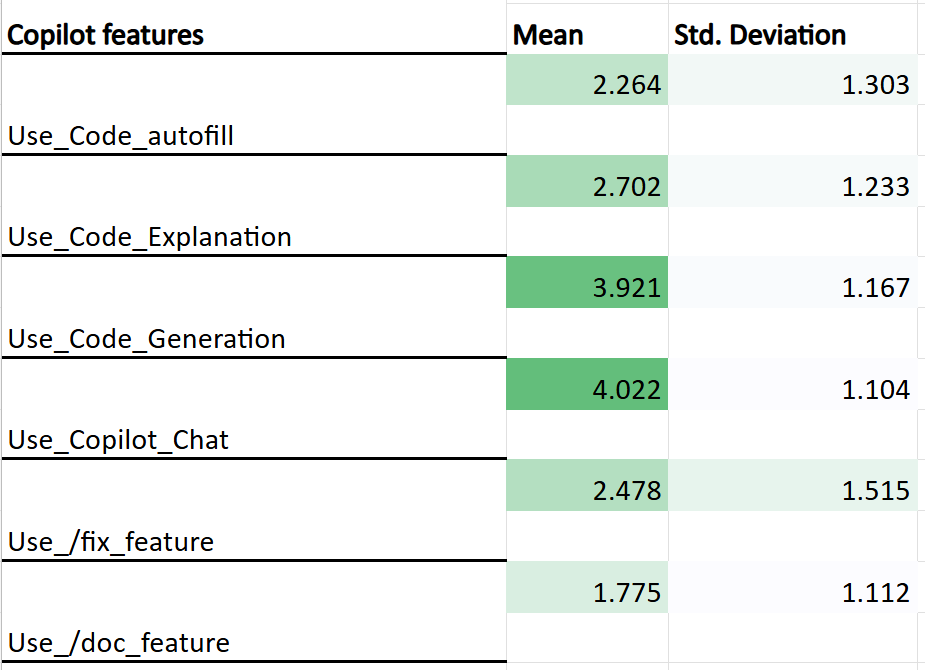} 
\caption{Figure shows the mean and standard deviation of the usage of GitHub Copilot features while working on an open-source contribution project. The response was recorded on a scale of 1 to 5 where 1 is "none at all" and 5 is "a great deal" }
\label{usage Copilot features}
\end{figure}

\begin{figure}[t]
\centering
\includegraphics[width=0.5\columnwidth]{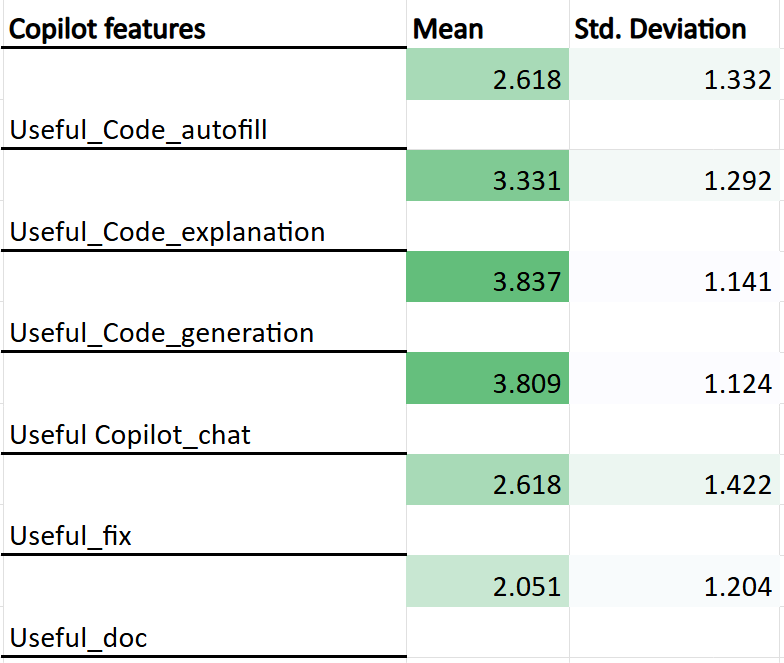} 
\caption{Figure shows the mean and standard deviation of how useful the participants found the GitHub Copilot features while working on an open-source contribution project. The response was recorded on a scale of 1 to 5 where 1 is "not at all useful" and 5 is "extremely useful"}
\label{usefulness Copilot features}
\end{figure}

\noindent
In response to participants' level of programming proficiency on a scale of 1 to 5, where 1 is no programming experience at all, and 5 is an advanced level in programming proficiency. participants reported a mean=2.33, SD=0.58. Showing students had some programming experience, although they were neither experts nor novices. Further, participants reported their familiarity with coding in C++ programming language; mean=3.84, SD=0.74. Familiarity with AI coding assistant was reported to be above median (mean=3.09, SD=1.04).
In terms of familiarity with LLM on a scale of 1 to 5, they reported a mean of 2.43 and an SD of 1.119.
Familiarity with AI coding assistant was reported to be above median (mean=3.09, SD=1.04). All the reported scores are also illustrated in Figure \ref{usage Copilot features}.
All familiarity questions used a scale of 1 to 5, where 1 is not at all and 5 is a great deal.
Further, participants responded to the question on their perceived usefulness of General LLMs such as ChatGPT and AI coding assistants. It was measured on a scale of 1 to 5, where 1 was not useful at all, and 5 was extremely useful. The reported scores are also illustrated in Figure \ref{usefulness Copilot features}.
Participants reported above median usefulness of Generic LLM such as ChatGPT as perceived by them (mean=3.51, SD=1.04). 
Participants also reported above median usefulness of AI coding assistant (mean=3.62, SD=1.05).

\subsection{GitHub Copilot Feature Usage and its Perceived Usefulness in the Open-Source Project Contribution}
\noindent
Overall, all the participants used some of the features offered by GitHub Copilot coding assistant. However, the usage of each GitHub Copilot feature varied.
Participants rate their usage of each GitHub Copilot feature on a scale of 1 to 5, where 2 is not at all and 5 is a great deal. Overall, the mean usage of Code autofill feature was 2.26 with SD= 1.30, mean of usage of Code explanation is 2.70 with as SD=1.23, the mean usage of "Code generation" was 3.92 with SD=1.16, mean usage of "Copilot chat" was 4.02 and SD=1.10, mean usage of "/fix" feature is 2.47 with SD=1.51, the mean usage of "/doc" feature is 1.77 with SD=1.11.
Participants also reported how useful they found each GitHub Copilot feature in completing the open-source project contribution task.
The self-reported scores were collected on a scale of 1 to 5, with 1 indicating not useful at all and 5 indicating extremely useful.
Participants reported somewhat above the median usefulness of the GitHub Copilot code autofill feature (mean=2.61, SD=1.33). 
Participants reported above median usefulness of the Code explanation feature (mean=3.33, SD=1.29).
Participants reported the code generation feature to be above the median in usefulness (mean=3.83, SD=1.14).
Above median scores were also reported for the usefulness of the Copilot chat feature (mean=3.80, SD=1.12).
Below median scores were reported for the Copilot/fix feature (mean=2.61, SD=1.42).
Participants also reported much below the median usefulness of the Copilot/doc feature (mean=2.05, 1.20).

\subsection{Difference in GitHub Copilot Usage and Self-Reported Scores Based on Gender, Prior Familiarity, and Programming Proficiency}
\noindent
\textbf{GitHub Copilot usage and perception differences based on Gender}
\noindent
\begin{figure}[t]
    \centering
    \makebox[0.8\textwidth][c]{%
        \begin{subfigure}{0.45\textwidth}
            \centering
            \includegraphics[width=\linewidth]{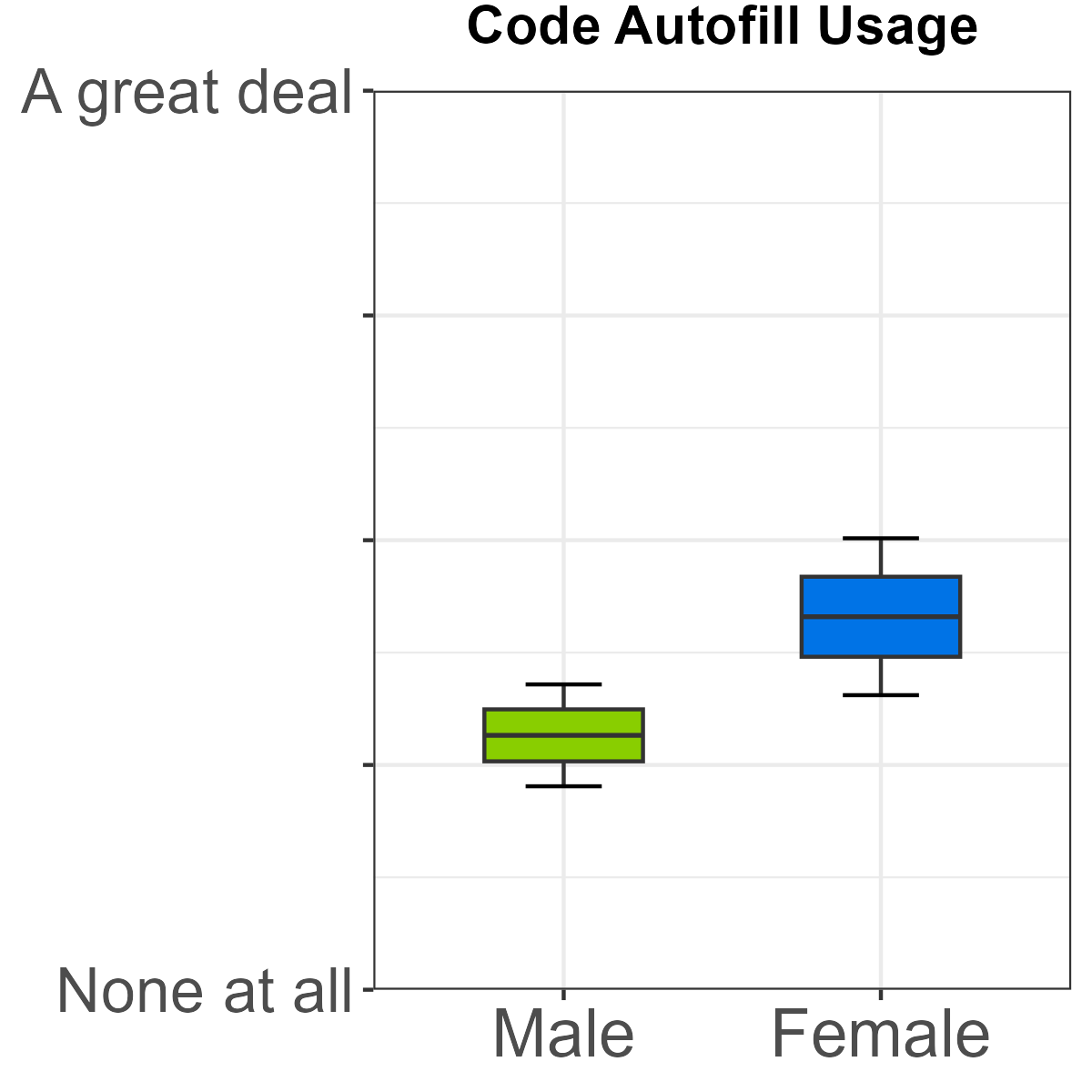}
            \label{fig:genderAutofillUse}
        \end{subfigure}%
        \hfill
        \begin{subfigure}{0.45\textwidth}
            \centering
            \includegraphics[width=\linewidth]{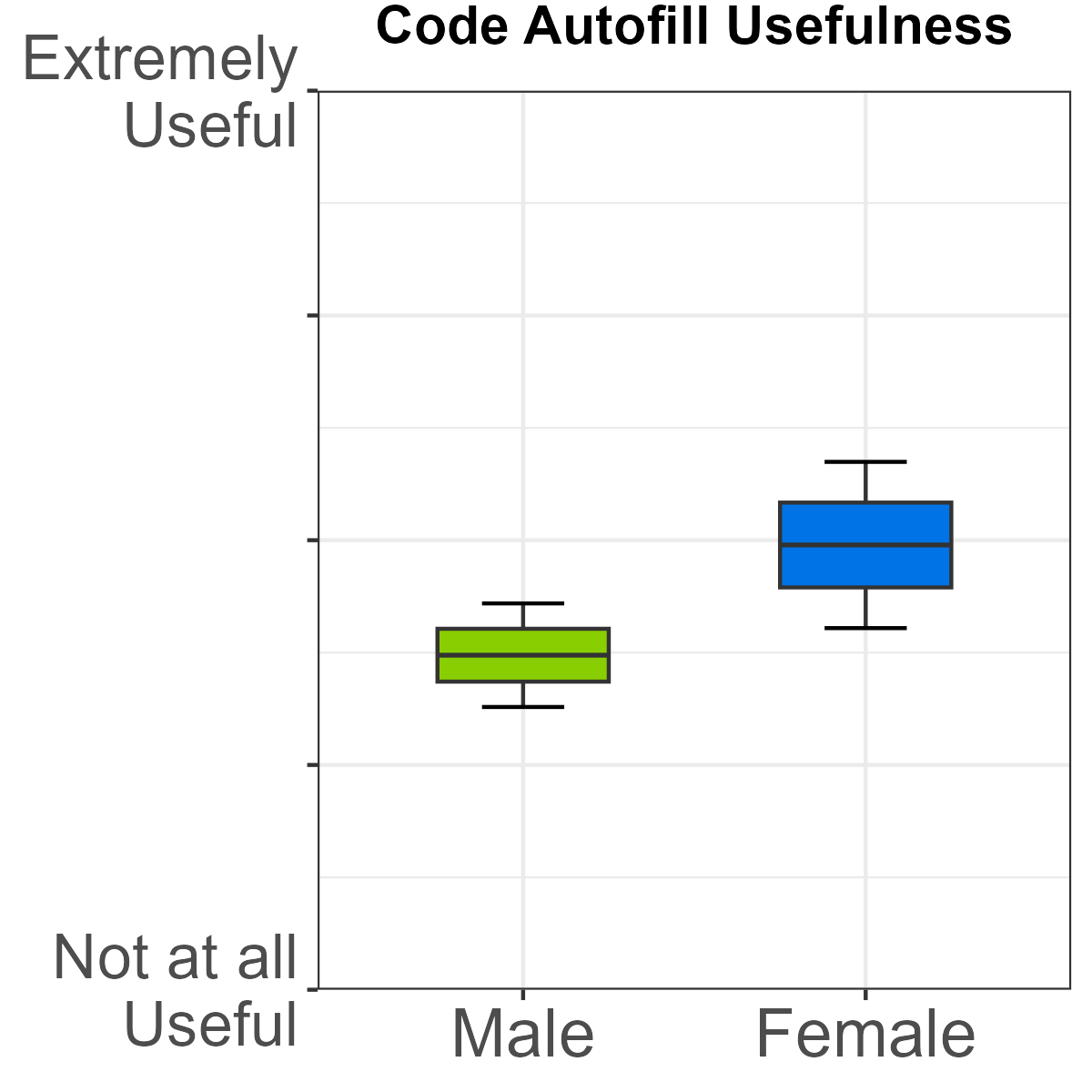}
            \label{fig:genderAutofillUseful}
        \end{subfigure}%
    }
    \caption{Female participants reported significantly higher usage and perceived usefulness of the code autofill feature than male participants.}
    \label{fig:genderAutofill}
\end{figure} 
Results of the Mann-Whitney U test show a near-significant difference between male and female participants in their familiarity with LLM (p=0.059). A small effect size was observed with r=.14. Male participants (mean=2.52, SD=1.13) reported higher familiarity with LLM as opposed to female participants (mean=2.15, SD=1.04).
There was also a significant difference in the perceived usefulness of the AI coding assistant based on gender (p=0.002). A small effect size was observed with r=.23. Male participants (mean=3.76, SD=1.04) reported significantly higher usefulness of AI coding assistant as opposed to female participants (mean=3.23, SD=1.02).



\begin{figure}[t]
    \centering
    \makebox[0.8\textwidth][c]{%
        \begin{subfigure}{0.45\textwidth}
            \centering
            \includegraphics[width=\linewidth]{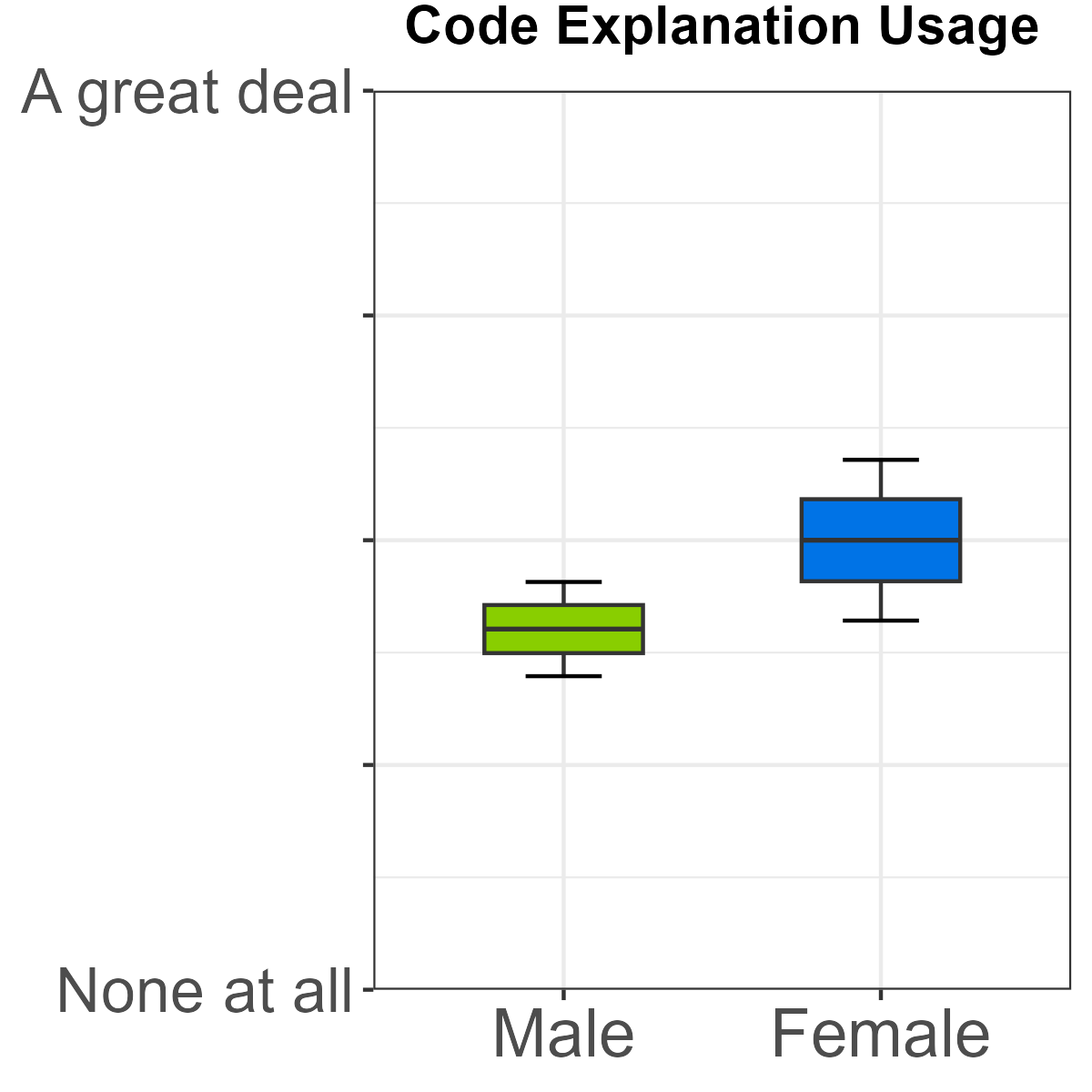}
            \label{fig:genderExplanationUse}
        \end{subfigure}%
        \hfill
        \begin{subfigure}{0.45\textwidth}
            \centering
            \includegraphics[width=\linewidth]{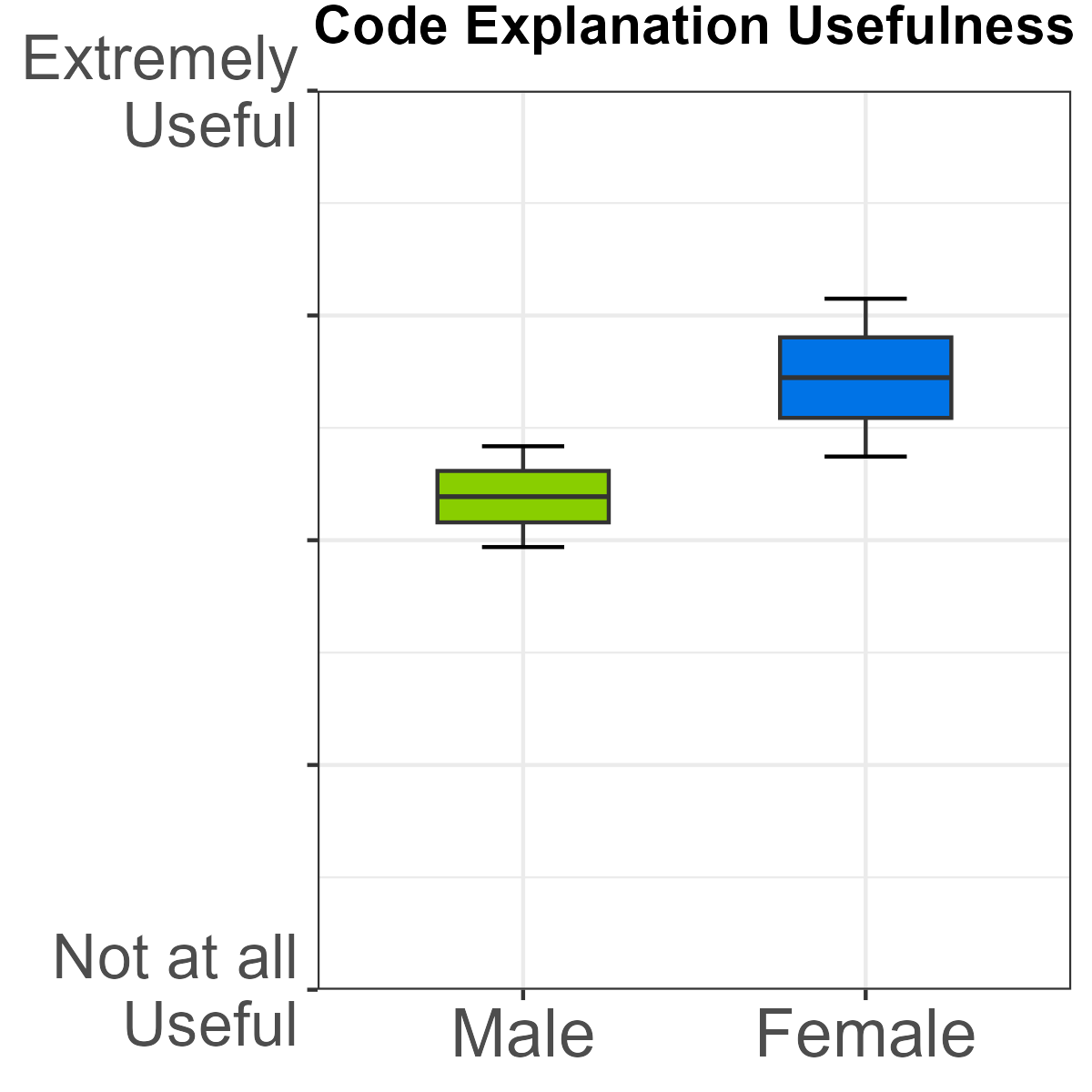}
            \label{fig:genderExplanationUseful}
        \end{subfigure}%
    }
    \caption{Female participants reported significantly higher usage and perceived usefulness of the code explanation feature than male participants.}
    \label{fig:genderExplanation}
\end{figure} 

\bigbreak
\noindent
Further significant differences were reported in the use of the code autofill feature by gender (p=0.008). 
Female participants (mean=2.66, SD=1.22) used code autofill significantly more than male participants (mean=2.13, SD=1.31). A small effect size with r=.20 was observed. This is illustrated in Figure \ref{fig:genderAutofill}.
Use of the code generation feature was also significantly different between males and females (p=0.02). Male participants (mean=4.01, SD=1.17) used code generation significantly more than females (mean=3.63, SD=1.13). A small effect size with r=.15 was observed.
A near-significant difference was observed between male and female usage of the code explanation feature (p=0.056), with female participants (mean=3.00, SD=1.25) using it more than males (mean=2.60, SD=1.21). A small size effect with r=.16 was observed. This is illustrated in Figure \ref{fig:genderExplanation}.
A near-significant difference was also observed in the usage of the/doc feature based on gender (p=0.056), with female participants (mean=1.95, SD=1.04) using it more than males (mean=1.72, SD=1.13). A small effect size (r = .10) was observed.
Further, the usefulness of the GitHub Copilot feature also varied by gender. A significant difference was observed between males and females in the reported usefulness of the code autofill feature. Female participants (mean=2.66, SD=1.22) found code autofill to be more useful than male participants (mean=2.13, SD=1.31). A small effect size (r = .20) was observed. This is also illustrated in Figure \ref{fig:genderAutofill}.
Significant difference was also observed in the perceived usefulness of the code explanation feature in the task between male and female participants. Females (mean=3.72, SD=1.22) reported significantly higher usefulness than males (mean=3.19, SD=1.30). A small effect size (r = .20) was observed. This is illustrated in Figure \ref{fig:genderExplanation}.
The usefulness of the code-generation feature also differed by gender. Male participants (mean=3.91, SD=1.18) reported significantly higher usefulness of code generation feature as opposed to female participants (mean=3.63, SD=1.00). A small effect size (r = .12) was observed.

\begin{figure}[t]
    \centering
    \makebox[0.8\textwidth][c]{%
        \begin{subfigure}{0.45\textwidth}
            \centering
            \includegraphics[width=\linewidth]{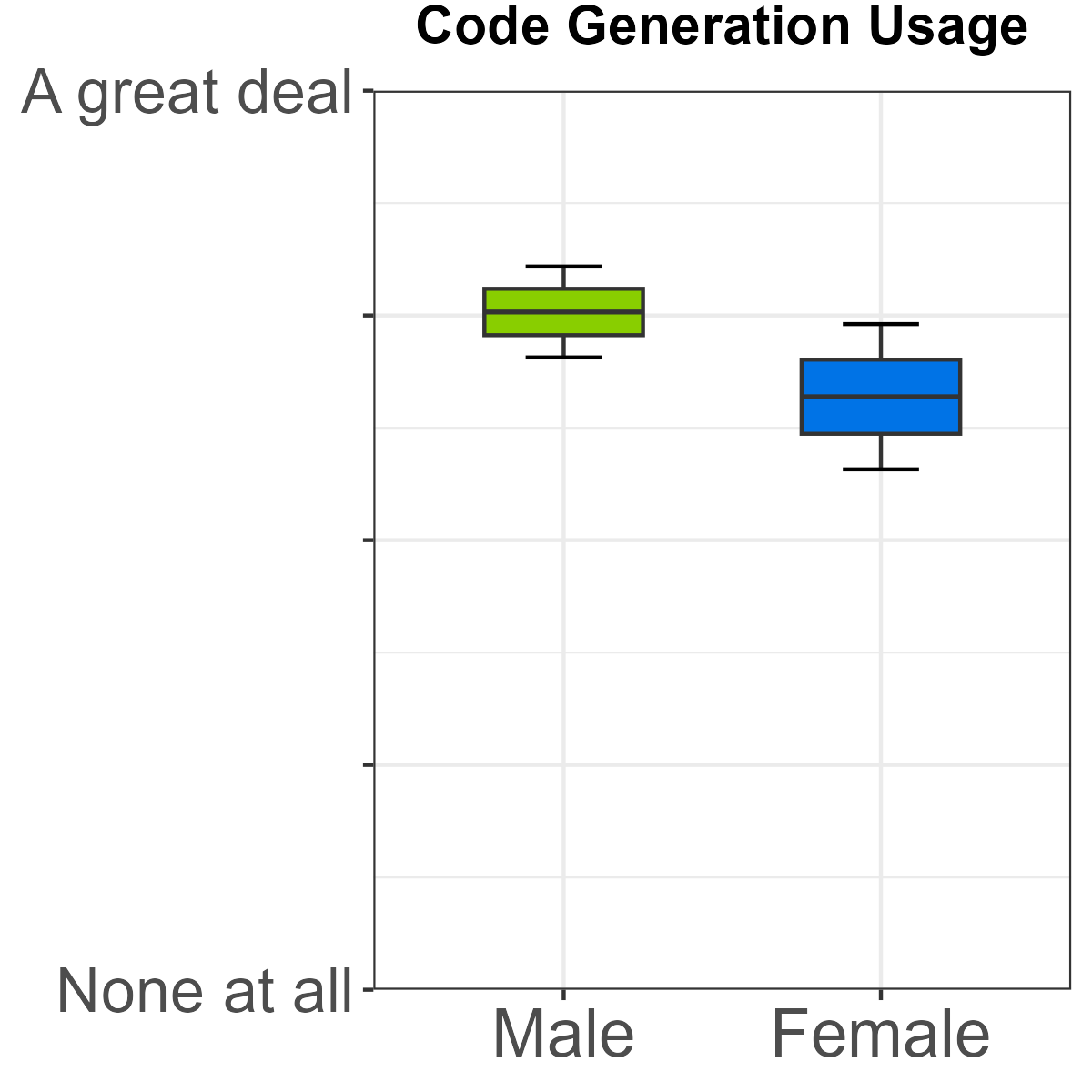}
            \label{fig:genderGenerationUse}
        \end{subfigure}%
        \hfill
        \begin{subfigure}{0.45\textwidth}
            \centering
            \includegraphics[width=\linewidth]{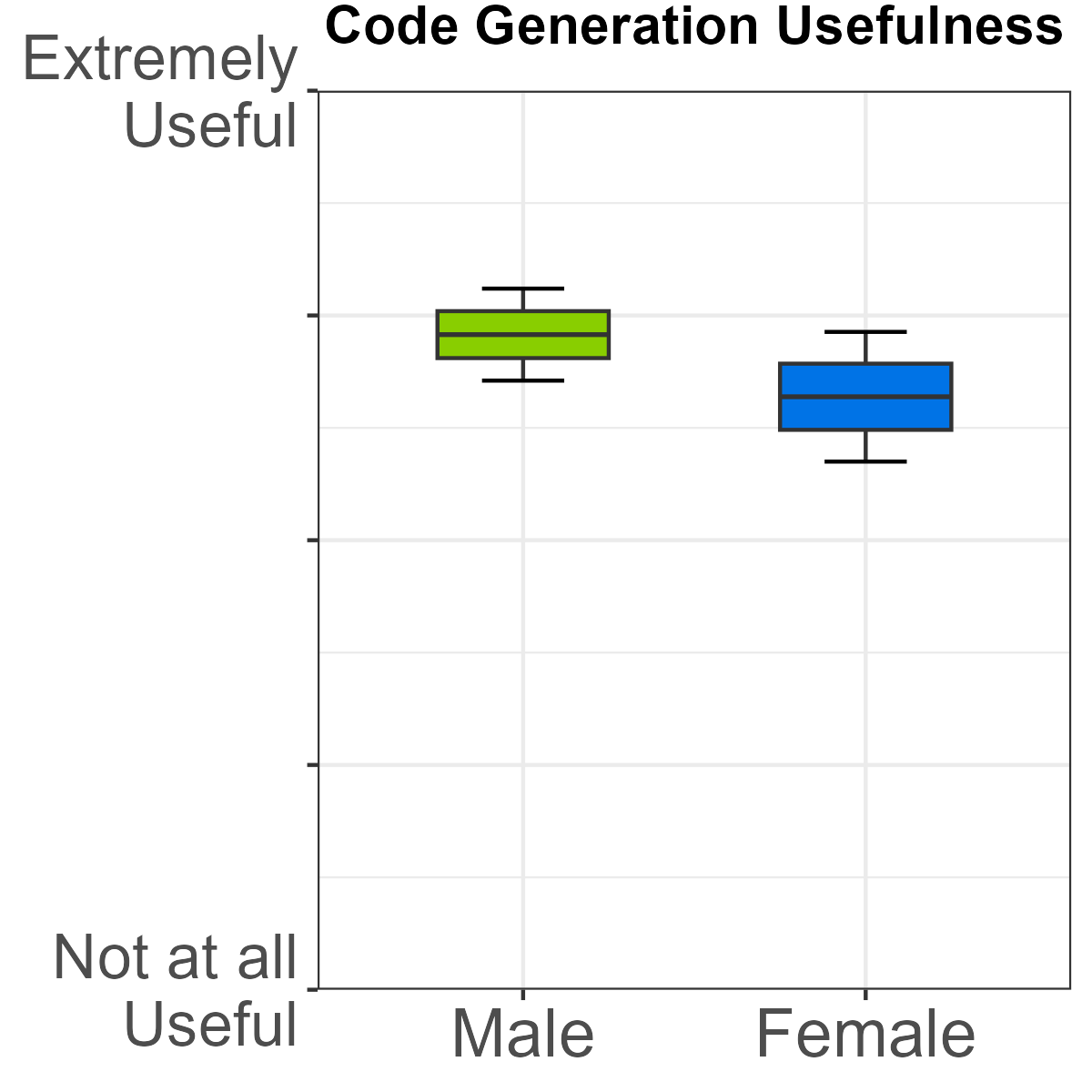}
            \label{fig:genderGenerationUseful}
        \end{subfigure}%
    }
    \caption{Male participants reported significantly higher usage and perceived usefulness of the code generation feature than female participants.}
    \label{fig:genderAutofill}
\end{figure} 

\bigbreak
\noindent
\textbf{GitHub Copilot usage and perception differences based on Programming proficiency.}
\noindent
Results of the Chi-square test show a significant difference in usage of GitHub Copilot chat feature based on participants' programming proficiency \(\chi ^{2}(8, N=178)=17.05,p=.03\)). It was observed that participants with low programming proficiency used the chat feature very little, but people with medium and high levels of programming proficiency used the chat feature significantly more.
Furthermore, participants perceived overall faith in GitHub Copilot differed significantly based on their programming proficiency \(\chi ^{2}(12, N=178)=22.95,p=.028\)). An effect size of r=.35 was observed. People with higher programming proficiency tend to have about median faith in GitHub Copilot, but people with somewhat programming proficiency tend to have somewhat higher faith.
Level of familiarity with C++, the major programming language used in the open-source project, impacted participants' perceived reliability on GitHub Copilot \(\chi^2 (18,N=178)=30.84,p=.03\). A medium effect size (r=.41) was observed. As 
The level of familiarity with C++, also impacted participants' perceived faith in GitHub Copilot \(\chi^2 (18,N=178)=31.51,p=.02\).A medium effect size (r=.41) was observed.

\bigbreak
\noindent
\textbf{GitHub Copilot usage and perception differences based on familiarity with AI coding assistant.}
\noindent
Results of the Chi-square test show a significant difference in overall trust in the GitHub Copilot feature based on participants' level of familiarity with the AI coding assistant \(\chi ^{2}(20,N=178)=38.20,p=.008\)). A medium-sized effect was observed (r=.45).
Furthermore, participants' overall perceived understandability of GitHub Copilot differed significantly based on their programming proficiency \(\chi ^{2}(24,N=178)=49.54,p=.002\)). A medium-sized effect was observed (r=.52).
Level of familiarity with AI coding assistant impacted participants' personal attachment with GitHub Copilot \(\chi^2 (24,N=178)=52.67,p<.001\). A medium-sized effect was observed (r=.54).
Further, the level of familiarity with AI coding assistant impacted participants' usage of GitHub Copilot /fix feature \(\chi ^2 (24,N=178)=52.67,p<.001\). A medium-sized effect (r=.54) was observed.

\section{Discussion}
\noindent
The study participants were the students who enrolled in the Software Engineering course and were enrolled in computing majors, so they had some familiarity with LLMs and AI coding assistants. They had above-median coding experience with C++ and below-median programming proficiency. Overall, they had some prior familiarity with the programming language C++ and AI coding assistants, but they were not experts and were not very confident in their programming skills.
We also found that the level of familiarity with AI coding assistants impacted their overall trust in GitHub Copilot. This is in line with the usual belief that being familiar with a tool is likely to impact your overall trust in it. The Level of familiarity with the AI coding assistant especially impacted trust constructs such as understandability and personal attachment. It is somewhat expected that people who have used AI coding assistants before and are familiar with them will have a better understanding of them. 

In terms of usage trends, we observed that the GitHub Copilot chat feature was used a lot, followed by the code generation feature. This observation is in line with prior work where the Chat feature was used the most by the upper division students \cite{10.1145/3641554.3701800}. Given that the chat feature allows users to ask any question tailored to their preferences and specific to their needs, it makes it a preferred feature. Further, given the popularity of Generative AI, where users interact through chat features, students likely felt more comfortable using this feature due to prior usage experience as reported in the familiarity questionnaire. Generally, students reported above median familiarity with Generic LLMs such as ChatGPT and with AI coding assistants. Highest perceived usefulness was also reported for the chat feature and the code generation feature. Possibly, the higher usefulness led to higher use of these features. The Copilot/doc was the least used feature. This made sense as students were primarily concerned with making the code contribution according to the assignment, and the coding standard and documentation were not prioritized. This indicates students' programming tendency of just doing the programming and not being invested in holistic software development. This result might change in other software projects where documentation is prioritized.

\bigbreak
\noindent
We also observed a significant impact of programming proficiency on the usage of GitHub Copilot chat features. This is in line with the expectation that students who know programming better are likely to chat and continue prompting till they get suitable code, while students who do not know programming well will prompt fewer times, as they would not be able to identify the fixes that the generated code needs, hence lesser prompting and chat interaction. Further, students who know programming well are likely to place appropriate faith in the AI coding assistant, while students with less programming proficiency are likely to over-rely or under-rely on the AI coding assistant. Literature has also indicated the possible risk of over-reliance on AI, especially with novices \cite{10.1145/3641554.3701800, pitts2025students, pitts2026trust}. 
We also observed differences in GitHub Copilot feature usage and overall trust towards it based on students' familiarity with AI coding assistants. This generally makes sense, and prior experience with and awareness of AI are likely to shape perceptions, preferences, and trust towards it. 

\bigbreak
\noindent
Interestingly, we found differences in the level of familiarity with LLM based on gender. Female students had near significantly lower familiarity with LLMs compared to male students. Further, we also found that gender played a significant role in the perceived usefulness of an AI coding assistant. Males had a higher perceived usefulness of the AI coding assistant compared to females. This can be somewhat related to recent observations of females having lesser knowledge about AI and its experience than their male counterparts \cite{cachero2025gender}. Lesser experience may lead to a different perception than usage with prior experience. This calls for special attention towards female AI training and awareness. A possible direction for future research. To mitigate gender-based differences in perception, introducing students to AI and systems understanding could be useful \cite{iddrisu2025gender}. Developing a better understanding and gaining exposure to AI can help bridge the gender gaps in prior AI experience. This may help in creating appropriate student reliance through system understanding \cite{pitts2025students}.

\section{Conclusion}
\noindent
Our investigation and results help build our understanding of how the usage of AI-based coding assistance exists among computer science students. 
We explore the usage and perception of GitHub Copilot, an AI coding assistant in real-world unconstrained settings. 
Students were free to use the GitHub Copilot features as they desired, and all the students used GitHub Copilot features at least once. Further, they used some features more than others and found some features to be more useful than others. This shows that students who have some level of programming experience, although not an expert, tend to use AI for code generation. Further investigation of the acceptance of AI responses and how students evaluate the generated code and accept or reject it, is a future work direction. This will also help in understanding reliance on AI \cite{amoozadeh2024student}. Overall, we observed that students used the code generation and chat feature the most.
 
While we are exploring and allowing students to use AI, we should also be cautious about students' learning in the process, especially when educational institutions are observing an increase in AI \cite{StudentGenerativeAISurvey2025_2025}. Heavy reliance and usage of GitHub Copilot may have a negative impact on their learning \cite{vaithilingam2022expectation}, while it might allow them to approach the problems that they would not have with the AI assistance. 

We also made some interesting observations on gender differences in AI usage and perception. This is an emerging area of research where focus needs to be placed.
Overall, our investigation has multiple potential impacts. This understanding will help computer science, especially software engineering and programming, instructors understand how usage may differ across the student population and, therefore, design assignments and projects that put all students on a level playing field. Instructors can design coding assistance with chat and code-generation features, which students find most useful. This will also help us understand how different students use AI differently and how much AI helps them, and subsequently, the quality of their work.

\section{Future Work}
\noindent
In their future, we will explore the GitHub Copilot usage pattern among students and what scenarios they reach out to the AI coding assistant for help. A direction for future investigators is to understand how students approach using an AI coding assistant, such as GitHub Copilot, to work on projects and languages they are not familiar with. Research in this direction will help understand the space of coding with an AI coding assistant, which may help prepare for future work in an evolving AI-rich work environment.

\section{Limitations}
\noindent
One of the limitations of the study is that we gathered and analyzed the self-reported usage of GitHub Copilot features and the perception towards them. This limitation was inherent in the nature of the task (open-source project contribution), being a take-home assignment. However, we believe our investigation and results are still useful as they build our understanding of how the space usage of AI-based coding assistance exists among computer science students. 
Another limitation of the study was that the groups that we compared were not of equal size. The class did not equal number of male and female students. Consequently, the study participant pool did not have an equal number of male and female participants. Further, the group sizes based on programming proficiency, prior familiarity with programming, and AI were not equal. However, it was unavoidable due to the study being conducted in a real classroom using a quasi-experimental methodology.

\section{Acknowledgments}
\noindent
We would like to thank all the participants in the study for agreeing to take part.

\bibliography{ref}

@article{pitts2026trust,
  title={Trust and Reliance on AI in Education: AI Literacy and Need for Cognition as Moderators},
  author={Pitts, Griffin and Rani, Neha and Mildort, Weedguet},
  journal={arXiv preprint arXiv:2604.01114},
  year={2026}
}

@article{rani2025can,
  title={Can AI support student engagement in classroom activities in higher education?},
  author={Rani, Neha and Majumder, Sharan and Bhardwaj, Ishan and Garcia, Pedro Guillermo Feijoo},
  journal={arXiv preprint arXiv:2506.18941},
  year={2025}
}

@article{xiao2024generative,
  title={Generative AI for pull request descriptions: Adoption, impact, and developer interventions},
  author={Xiao, Tao and Hata, Hideaki and Treude, Christoph and Matsumoto, Kenichi},
  journal={Proceedings of the ACM on Software Engineering},
  volume={1},
  number={FSE},
  pages={1043--1065},
  year={2024},
  publisher={ACM New York, NY, USA}
}

@misc{stackoverflow2025survey,
  title        = {Stack Overflow Developer Survey 2025},
  author       = {{Stack Overflow}},
  year         = {2025},
  howpublished = {\url{https://survey.stackoverflow.co/2025/}},
  note         = {Accessed: 2026-01-12}
}

@article{zhou2025exploring,
  title={Exploring the problems, their causes and solutions of AI pair programming: A study on GitHub and Stack Overflow},
  author={Zhou, Xiyu and Liang, Peng and Zhang, Beiqi and Li, Zengyang and Ahmad, Aakash and Shahin, Mojtaba and Waseem, Muhammad},
  journal={Journal of Systems and Software},
  volume={219},
  pages={112204},
  year={2025},
  publisher={Elsevier}
}

@inproceedings{madsen2000measuring,
  title={Measuring human-computer trust},
  author={Madsen, Maria and Gregor, Shirley},
  booktitle={11th australasian conference on information systems},
  volume={53},
  pages={6--8},
  year={2000},
  organization={Citeseer}
}

@inproceedings{vaithilingam2022expectation,
  title={Expectation vs. experience: Evaluating the usability of code generation tools powered by large language models},
  author={Vaithilingam, Priyan and Zhang, Tianyi and Glassman, Elena L},
  booktitle={Chi conference on human factors in computing systems extended abstracts},
  pages={1--7},
  year={2022}
}

@inproceedings{begel2008struggles,
  title={Struggles of new college graduates in their first software development job},
  author={Begel, Andrew and Simon, Beth},
  booktitle={Proceedings of the 39th SIGCSE technical symposium on Computer science education},
  pages={226--230},
  year={2008},
}

@article{craig2018listening,
  title={Listening to early career software developers},
  author={Craig, Michelle and Conrad, Phill and Lynch, Dylan and Lee, Natasha and Anthony, Laura},
  journal={Journal of Computing Sciences in Colleges},
  volume={33},
  number={4},
  pages={138--149},
  year={2018},
  publisher={Consortium for Computing Sciences in Colleges}
}

@inproceedings{wermelinger2023using,
  title={Using github copilot to solve simple programming problems},
  author={Wermelinger, Michel},
  booktitle={Proceedings of the 54th ACM Technical Symposium on Computer Science Education V. 1},
  pages={172--178},
  year={2023}
}

@inproceedings{imai2022github,
  title={Is github copilot a substitute for human pair-programming? an empirical study},
  author={Imai, Saki},
  booktitle={Proceedings of the ACM/IEEE 44th International Conference on Software Engineering: Companion Proceedings},
  pages={319--321},
  year={2022}
}

@article{zhang2023practices,
  title={Practices and challenges of using github copilot: An empirical study},
  author={Zhang, Beiqi and Liang, Peng and Zhou, Xiyu and Ahmad, Aakash and Waseem, Muhammad},
  journal={arXiv preprint arXiv:2303.08733},
  year={2023}
}

@article{xiao2025self,
  title={Self-admitted GenAI usage in open-source software},
  author={Xiao, Tao and Fan, Youmei and Calefato, Fabio and Treude, Christoph and Kula, Raula Gaikovina and Hata, Hideaki and Baltes, Sebastian},
  journal={arXiv preprint arXiv:2507.10422},
  year={2025}
}

@article{song2024impact,
  title={The impact of generative AI on collaborative open-source software development: Evidence from GitHub Copilot},
  author={Song, Fangchen and Agarwal, Ashish and Wen, Wen},
  journal={arXiv preprint arXiv:2410.02091},
  year={2024}
}

@article{stray2025developer,
  title={Developer Productivity With and Without GitHub Copilot: A Longitudinal Mixed-Methods Case Study},
  author={Stray, Viktoria and Brandtz{\ae}g, Elias Goldmann and Wivestad, Viggo Tellefsen and Barbala, Astri and Moe, Nils Brede},
  journal={arXiv preprint arXiv:2509.20353},
  year={2025}
}

@inproceedings{ziegler2022productivity,
  title={Productivity assessment of neural code completion},
  author={Ziegler, Albert and Kalliamvakou, Eirini and Li, X Alice and Rice, Andrew and Rifkin, Devon and Simister, Shawn and Sittampalam, Ganesh and Aftandilian, Edward},
  booktitle={Proceedings of the 6th ACM SIGPLAN International Symposium on Machine Programming},
  pages={21--29},
  year={2022}
}

@inproceedings{sandoval2023lost,
  title={Lost at c: A user study on the security implications of large language model code assistants},
  author={Sandoval, Gustavo and Pearce, Hammond and Nys, Teo and Karri, Ramesh and Garg, Siddharth and Dolan-Gavitt, Brendan},
  booktitle={32nd USENIX Security Symposium (USENIX Security 23)},
  pages={2205--2222},
  year={2023}
}

@article{asare2023github,
  title={Is github’s copilot as bad as humans at introducing vulnerabilities in code?},
  author={Asare, Owura and Nagappan, Meiyappan and Asokan, Nirmal},
  journal={Empirical Software Engineering},
  volume={28},
  number={6},
  pages={129},
  year={2023},
  publisher={Springer}
}

@inproceedings{amoozadeh2024student,
  title={Student-ai interaction: A case study of CS1 students},
  author={Amoozadeh, Matin and Nam, Daye and Prol, Daniel and Alfageeh, Ali and Prather, James and Hilton, Michael and Srinivasa Ragavan, Sruti and Alipour, Amin},
  booktitle={Proceedings of the 24th Koli Calling International Conference on Computing Education Research},
  pages={1--13},
  year={2024}
}

@incollection{margulieux2024self,
  title={Self-regulation, self-efficacy, and fear of failure interactions with how novices use llms to solve programming problems},
  author={Margulieux, Lauren E and Prather, James and Reeves, Brent N and Becker, Brett A and Cetin Uzun, Gozde and Loksa, Dastyni and Leinonen, Juho and Denny, Paul},
  booktitle={Proceedings of the 2024 on Innovation and Technology in Computer Science Education V. 1},
  pages={276--282},
  year={2024}
}

@inproceedings{shihab2025effects,
  title={The Effects of GitHub Copilot on Computing Students' Programming Effectiveness, Efficiency, and Processes in Brownfield Coding Tasks},
  author={Shihab, Md Istiak Hossain and Hundhausen, Christopher and Tariq, Ahsun and Haque, Summit and Qiao, Yunhan and Mulanda, Brian Wise},
  booktitle={Proceedings of the 2025 ACM Conference on International Computing Education Research V. 1},
  pages={407--420},
  year={2025}
}

@article{fan2025impact,
  title={The impact of AI-assisted pair programming on student motivation, programming anxiety, collaborative learning, and programming performance: a comparative study with traditional pair programming and individual approaches},
  author={Fan, Guangrui and Liu, Dandan and Zhang, Rui and Pan, Lihu},
  journal={International Journal of STEM Education},
  volume={12},
  number={1},
  pages={16},
  year={2025},
  publisher={Springer}
}

@misc{StudentGenerativeAISurvey2025_2025,
url={https://www.hepi.ac.uk/reports/student-generative-ai-survey-2025/}, 
abstractNote={On 26 February, HEPI and Kortext will publish the Student Generative AI Survey 2025, by Josh Freeman, with a Foreword by Professor Janice Kay CBE. Based on a survey of 1,041 students conducted by Savanta, the report shows an unprecedented increase in the use of generative AI tools among undergraduate students from the rates recorded […]}, 
journal={HEPI}, 
year={2025}, 
month=feb, 
language={en-GB} 
}

@inproceedings{Li2025AITeammates,
author={Li, Hao and Zhang, Haoxiang and Hassan, Ahmed E.},
title={The Rise of AI Teammates in Software Engineering (SE) 3.0: How Autonomous Coding Agents Are Reshaping Software Engineering},
booktitle={Proceedings of the XX},
year={2025},
publisher={ACM},
address={New York, NY, USA},
pages={1--22}
}

@online{dohmke2023economic,
author={Dohmke, Thomas},
title={The economic impact of the AI-powered developer lifecycle and lessons from GitHub Copilot},
year={2023},
month={June 27},
url={https://github.blog/news-insights/research/the-economic-impact-of-the-ai-powered-developer-lifecycle-and-lessons-from-github-copilot/},
organization={GitHub Blog}
}

@inproceedings{pitts2025students,
  title={Students’ reliance on ai in higher education: identifying contributing factors},
  author={Pitts, Griffin and Rani, Neha and Mildort, Weedguet and Cook, Eva-Marie},
  booktitle={International Conference on Human-Computer Interaction},
  pages={86--97},
  year={2025},
  organization={Springer}
}

@Article{su15075614,
AUTHOR = {Kooli, Chokri},
TITLE = {Chatbots in Education and Research: A Critical Examination of Ethical Implications and Solutions},
JOURNAL = {Sustainability},
VOLUME = {15},
YEAR = {2023},
NUMBER = {7},
ARTICLE-NUMBER = {5614},
URL = {https://www.mdpi.com/2071-1050/15/7/5614},
ISSN = {2071-1050},
DOI = {10.3390/su15075614}
}

@inproceedings{ghimire2024coding,
  title={Coding with ai: How are tools like chatgpt being used by students in foundational programming courses},
  author={Ghimire, Aashish and Edwards, John},
  booktitle={International Conference on Artificial Intelligence in Education},
  pages={259--267},
  year={2024},
  organization={Springer}
}

@inproceedings{10.1145/3643991.3645078,
author = {Mohamed, Suad and Parvin, Abdullah and Parra, Esteban},
title = {Chatting with AI: Deciphering Developer Conversations with ChatGPT},
year = {2024},
isbn = {9798400705878},
publisher = {Association for Computing Machinery},
address = {New York, NY, USA},
url = {https://doi.org/10.1145/3643991.3645078},
doi = {10.1145/3643991.3645078},
booktitle = {Proceedings of the 21st International Conference on Mining Software Repositories},
pages = {187–191},
numpages = {5},
location = {Lisbon, Portugal},
series = {MSR '24}
}

@inproceedings{10.1145/3641554.3701800,
author = {Shah, Anshul and Chernova, Anya and Tomson, Elena and Porter, Leo and Griswold, William G. and Soosai Raj, Adalbert Gerald},
title = {Students' Use of GitHub Copilot for Working with Large Code Bases},
year = {2025},
isbn = {9798400705311},
publisher = {Association for Computing Machinery},
address = {New York, NY, USA},
url = {https://doi.org/10.1145/3641554.3701800},
doi = {10.1145/3641554.3701800},
booktitle = {Proceedings of the 56th ACM Technical Symposium on Computer Science Education V. 1},
pages = {1050–1056},
numpages = {7},
location = {Pittsburgh, PA, USA},
series = {SIGCSETS 2025}
}

@inproceedings{radermacher2014investigating,
  title={Investigating the skill gap between graduating students and industry expectations},
  author={Radermacher, Alex and Walia, Gursimran and Knudson, Dean},
  booktitle={Companion Proceedings of the 36th international conference on software engineering},
  pages={291--300},
  year={2014}
}

@article{iddrisu2025gender,
  title={Gender differences in the adoption, usage, and perceived effectiveness of AI writing tools: a study among university for development studies students},
  author={Iddrisu, Hassan Mubarik and Iddrisu, Simon Alhassan and Aminu, Bashiru},
  journal={International Journal of Educational Innovation and Research},
  volume={4},
  number={1},
  pages={110--111},
  year={2025}
}

@article{cachero2025gender,
  title={Gender Bias in Self-Perception of Artificial Intelligence Knowledge, Impact, and Support Among Higher Education Students: An Observational Study},
  author={Cachero, Cristina and Tom{\'a}s, David and Pujol, Francisco A and others},
  year={2025},
  publisher={Association for Computing Machinery (ACM)}
}

\appendix
\section{Appendix}
\subsection{Trust Statements}
\label{sec:trustAppendix}
\noindent
This appendix contains the statements for each of the trust constructs used in the post-study questionnaire. 
The constructs and their statements are taken from a human-computer trust scale developed by Madsen and Gregor~\cite{madsen2000measuring}, modified to replace "the system" with "Copilot".
Participants were asked to rate their agreement with each of the following statements on a 7-point Likert scale, where 1 is "Strongly Disagree" and 7 is "Strongly Agree":

\subsubsection{Perceived Reliability}
\begin{itemize}
    \item Copilot always provides the advice I require to make my decision.
    \item Copilot performs reliably.
    \item Copilot responds the same way under the same conditions at different times.
    \item I can rely on Copilot to function properly.
    \item Copilot analyzes problems consistently.
\end{itemize}
\subsubsection{Perceived Technical Competence}
\begin{itemize}
    \item Copilot uses appropriate methods to reach decisions.
    \item Copilot has sound knowledge about this type of problem built into it.
    \item The advice Copilot produces is as good as that which a highly competent person could produce.
    \item Copilot correctly uses the information I enter.
    \item Copilot makes use of all the knowledge and information available to it to produce its solution to the problem.
\end{itemize}
\subsubsection{Perceived Understandability}
\begin{itemize}
    \item I know what will happen the next time I use Copilot because I understand how it behaves.
    \item I understand how Copilot will assist me with decisions I have to make.
    \item Although I may not know how Copilot works, I know how to use it to make decisions about the problem.
    \item It is easy to follow what Copilot does.
    \item I recognize what I should do to get the advice I need from Copilot the next time I use it.
\end{itemize}
\subsubsection{Faith}
\begin{itemize}
    \item I believe advice from Copilot even when I don't know for certain that it is correct.
    \item When I am uncertain about a decision I believe Copilot rather than myself.
    \item If I am not sure about a decision, I have faith that Copilot will proide the best solution.
    \item When Copilot gives unusual advice I am confident that the advice is correct.
    \item Even if I have no reason to expect Copilot will be able to solve a difficult problem, I still feel certain that it will.
\end{itemize}
\subsubsection{Personal Attachment}
\begin{itemize}
    \item I would feel a sense of loss if Copilot was unavailable and I could no longer use it.
    \item I feel a sense of attachment to using Copilot.
    \item I find Copilot suitable to my style of decision making.
    \item I like using Copilot for decision making.
    \item I have a personal preference for making decisions with Copilot.
\end{itemize}
\end{document}